# The Propagation of Photons in the Dilute Ionized Gas

Yijia Zheng

**National Astronomical Observatory, Chinese Academy of Sciences, Beijing, 100012, China**

**Abstract**

The dilute ionized gas is very popular in the Universe. Usually only the Compton interactions, e.g., the Sunyaev-Zel'dovich effect, were considered while photons propagated in this medium. In this paper the 'soft-photon process' is considered. Due to the soft photons emitted during the propagation of a photon in the dilute ionized gas, the main photon (propagating in the original direction) will be redshifted. The formula to calculate this redshift is derived.

## 1. Introduction

The dilute ionized gas is very popular in the Universe. It lies in the corona of stars and Warm-Hot Intergalactic Medium (WHIM). Usually only the Compton interactions, e.g., the Sunyaev-Zel'dovich effect, were considered while photons propagated in this medium. In this paper the 'soft-photon process' is considered. Due to the soft photons emitted during the propagation of a photon in the dilute ionized gas, the main photon (propagating in the original direction) will be redshifted. Because this redshift is very small, it is very hard to detect and has long been neglected. However, the soft-photon process has very important significance in astrophysics. The soft-photon process belongs to a reasonable 'tired light' theory.

In Section 2 the soft-photon process caused by the interaction between the photon and electron during the propagation process is described, and the formula to calculate the redshift is derived.

## 2. The propagation of photons in dilute ionized gas

The dilute ionized gas in the Universe mainly consists of ionized hydrogen. Therefore the dilute ionized gas contains a lot of free electrons. While photons propagate in this dilute ionized gas, the free electron in the ionized gas can absorb photons with any frequencies and jump from a lower energy level to a higher one. The free electron can also emit photons with any frequencies and jump from a higher energy level to a lower one. The emission process has two different types: the spontaneous emission and stimulated emission. The scattering process corresponds to the absorption and spontaneous emission process. The propagation process corresponds to the absorption and stimulated emission process. The Bremsstrahlung process corresponds to the spontaneous emission process with no absorption. The probability of the spontaneous emission can be calculated using the quantum electrodynamics theory. The probability of the stimulated emission is usually derived using the method developed by Einstein (Einstein 1917, Ginzburg 1970).

Let us consider a thin screen of the dilute ionized gas. The electron density in the screen and the thickness of the screen are denoted by $n_e$ and $dl = n_e^{-1/3}$, respectively. Consider a beam of photons incident perpendicularly to an area $ds$ of this screen. The total number of electrons that the beam of photons encounters is $N_e = n_e ds dl$. When the incident beam encounters an electron, due to the Compton scattering process a small fraction of incident photons will be scattered into other directions. However, most of the incident photons will propagate in the original direction (propagated photons). The number of the scattered photons is proportional to the Compton scattering cross section $\sigma_c$. In this case the number of propagated photons is proportional to ($ds - N_e \sigma_c$). Because in the dilute ionized gas the electron density $n_e$ is very small, $ds = N_e n_e^{(-2/3)}$ will be much larger than the Compton scattering cross section $N_e \sigma_c$. For example, even at the bottom of the solar corona, $n_e \approx 10^{12}$ /cm$^3$, the ratio of the probability of the propagate process to that of the Compton scattering process is approximately

$$\frac{\left(N_e n_e^{-2/3} - N_e \sigma_c\right)}{N_e \sigma_c} \approx \frac{n_e^{-2/3}}{\sigma_c} \approx 10^{16}$$

One may argue that there is no interaction between the propagated photons and the free electrons. Actually, according to the quantum theory the interaction between the photons and the free electrons always exists. The free electrons will absorb any photons with any frequencies and then re-emit photons. The propagation process includes the photon absorption and the subsequent stimulated photon re-emission in the original direction. During the propagation process a free electron in the dilute plasma absorbs an incident photon; the free electron will jump up from a lower energy level 1 to a higher energy level 2 and then stay in the energy level 2 for a while. In this period the free electron and absorbed photon are in an intermediate state. After a finite duration $\Delta t$, the free electron will be stimulated to emit a photon propagating in the original direction. Then, the free electron will drop down from the higher energy level 2 back to the lower energy level 1.

Because the propagation speed of photons is $c$ (the speed of light), the photon beam will take duration of $\Delta t = n_e^{-1/3} c^{-1}$ to pass through the screen with the thickness $dl = n_e^{-1/3}$. During this period the propagated photons are absorbed by the electron, and the electron stays in an intermediate state. Because of the interaction between the electron and other charged particles in the ionized gas, the momentum of the electron will be changed all the time. According to the semi-classical theory of radiation some soft photons will be emitted due to the momentum change of the electron. Because of the emission of the soft photons, the free electron (in the intermediate state) will drop from the energy level 2 to an energy level 2′ that is a little lower than the energy level 2. At the end of the intermediate state, it is the level 2′ from which the free electron jumps back to the energy level 1. Therefore the propagated photon will be redshifted relatively to the incident photon. The probability of the soft photons emitted by the electron is (Heitler 1954, Gould 1979)

$$F(P) = \frac{2e^2}{3\pi \hbar c} \frac{(\Delta P)^2}{E^2} \frac{dk_r}{k_r}$$

where $e$ is the charge of an electron, $\hbar$ the Planck constant, $c$ the speed of light, $\Delta P$ the momentum change of the free electron, $E$ the energy of the free electron, $k_r$ the energy of the soft photons emitted by the free electron, and $dk_r$ the energy range of the soft photons emitted.

In the propagation process due to the soft photon emission, the propagated photons will lose a small fraction of their energy. For each interaction between the propagated photon and an electron, the total energy loss is

$$-dk_1 = \int_0^k k_r F(P) = k \frac{2e^2}{3\pi \hbar c} \frac{(\Delta P)^2}{E^2} \qquad (1)$$

According to Spitzer (1956), in the dilute ionized gas, due to the affects of other charged particles, the velocity change of an electron can be calculated (see the appendix for detailed derivation). On the average, in the duration $\Delta t = n_e^{-1/3} c^{-1}$, due to the affects of other charged particles, the momentum change of the free electron is

$$(\Delta P)^2 = \frac{8\pi . e^4 n_e^{\frac{2}{3}} c \ln \Lambda}{V_e} \qquad (2)$$

where

$$\Lambda = \begin{cases} \dfrac{3}{2 z_e z_i e^3}\left(\dfrac{k^3 T^3}{\pi . n_e}\right)^{\frac{1}{2}} & \text{..............................} for......T < 4.2 \times 10^5 .°K \\ \dfrac{3}{2 z_e z_i e^3}\left(\dfrac{k^3 T^3}{\pi . n_e}\right)^{\frac{1}{2}}\left(\dfrac{4.2 \times 10^5}{T}\right)^{\frac{1}{2}} & \text{............} for.....T > 4.2 \times 10^5 .°K \end{cases}$$

Because

$$E = m_e c^2, \qquad \sigma_c = \frac{8\pi}{3}\left(\frac{e^2}{m_e c^2}\right)^2,$$

$$-dk_1 = k \frac{2e^2}{\pi . \hbar} \sigma_c \frac{\ln \Lambda}{V_e} n_e^{\frac{2}{3}} \qquad (3)$$

if the thickness of the thin screen is $dr$, then the total energy loss is

$$-dk = -dk_1\left(\frac{dr}{dl}\right) = k \frac{2e^2}{\pi . \hbar} \sigma_c \frac{\ln \Lambda}{V_e} n_e dr \qquad (4)$$

After the photon beam has propagated a distance R, the ratio of the total energy loss to the energy of the incident photon is

$$\int_{k_o}^{k} \frac{-dk}{k} = \frac{2e^2}{\pi . \hbar} \sigma_c \int_{0}^{R} \frac{\ln \Lambda}{V_e} n_e dr \qquad (5)$$

Because the redshift $Z$ is defined as

$$Z = \frac{v_o - v}{v} = \frac{k_o - k}{k}$$

the formula to calculate the redshift of the propagated photon is

$$\ln(1+Z) = \frac{2e^2}{\pi . \hbar} \sigma_c \int_{0}^{R} \frac{\ln \Lambda}{V_e} n_e dr \qquad (6)$$

## 3. Observational evidence

Pioneer 6 was injected into solar orbit on 16 December 1965, and ever since has reliably transmitted to earth by S-band telemetry the results of its scientific experiments. Spectrograms of the radio signals from Pioneer 6 were taken by Goldstein (Goldstein 1969) as the spacecraft was occulted by the sun. The signals from Pioneer 6 were of extremely narrow-band (monochromatic). Pioneer 6 normally transmits telemetry signals composed of a spectrally pure carrier wave plus a set of modulation sidebands. Goldstein only observed the carrier wave nominally at 2295 MHz (2292MHz in Merat et al. 1974). The sidebands, separated from the carrier wave by multiples of 2 kHz, were removed by filtering. The 210-foot (64-m) antenna of the Jet Propulsion Laboratory was used. This remarkable antenna has a beam width of only 0.14 degree at 2300 MHz (S-band). Such a narrow beam width was able to discriminate between the signal of Pioneer 6 and the powerful radio noise emitted by the sun. Normally, the antenna-receiver system is extremely sensitive, having an equivalent noise temperature of only 25 $^o$k. As the apparent distance R made by Pioneer 6, the earth and the sun diminished (see Fig. 1), the system temperature rose, until at R≈4$R_\odot$ where the temperature was between 200$^o$k and 300$^o$k. The signal disappeared when R was less than 4$R_\odot$. Apparently, this was due to the effect of the corona, because the signal-to-noise ratio still had been adequate for detection at smaller apparent distance R. In order to compensate for the orbital velocities of Pioneer 6 and the earth, the receiver was tuned continuously according to an ephemeris. This automatic tuning was accurate to 0.05 Hz. Slight frequency instability remained due to the spacecraft oscillator. A slow drift of approximately 13.4 Hz per day (Merat et al. 1974), presumably of thermal origin, was observed. In addition, a more erratic drift of up to 1 or 2 Hz per 15 minutes was observed. The frequency band used in the observation, 100 Hz, was the same for all the spectrograms taken. A filter defined this bandwidth at the last stage of the receiver.

As the spacecraft was occulted by the solar corona, the spectral bandwidths increased slowly at first, and then very rapidly at R≈4$R_\odot$. In addition, six solar "events" produced marked increases of bandwidth lasting for several hours. The received signal power seemed unaffected by the solar corona. The center frequency of the signals showed a shift. The drift depended on the apparent distance R made by Pioneer 6, the earth and the sun (see Fig 1). Goldstein (1969) claimed that the frequency shift was the effect of the line of sight across the corona.

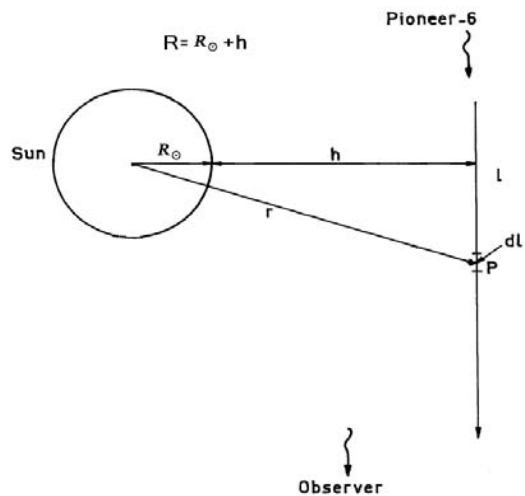

Fig. 1 The geometry of the problem

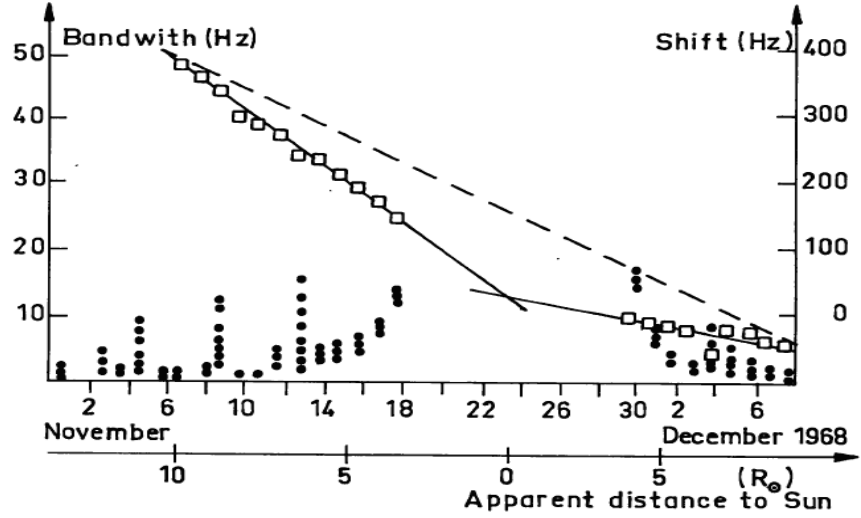

**Fig. 2** Observational results from Goldstein 1969. •: bandwidth; ▫: shift in center frequency. Dashed line represents a general 13.4 Hz drift per day during the experiment (reproduced from Fig. 1. of Chastel, et al. 1976).

If the redshift of the center frequency for the Pioneer 6 signal is caused by the solar corona when the signal passes through it, then using the smoothed corona electron density model given by Allen (1973), the redshift of the center frequency can be calculated numerically. Formula (6) can be simplified to formula (7) in the following because the calculated redshift Z is of the magnitude about $10^{-8}$ to $10^{-7}$ only.

$$Z = \frac{2e^2}{\pi \hbar} \sigma_c \int_0^L \frac{\ln \Lambda}{V_e} n_e(r) dr \qquad (7)$$

The variation of the factor FA=$\frac{2e^2}{\pi \hbar} \frac{\ln \Lambda}{V_e}$ in the solar atmosphere can be calculated, and the variation with the distance to center of the sun is shown in Fig. 3.

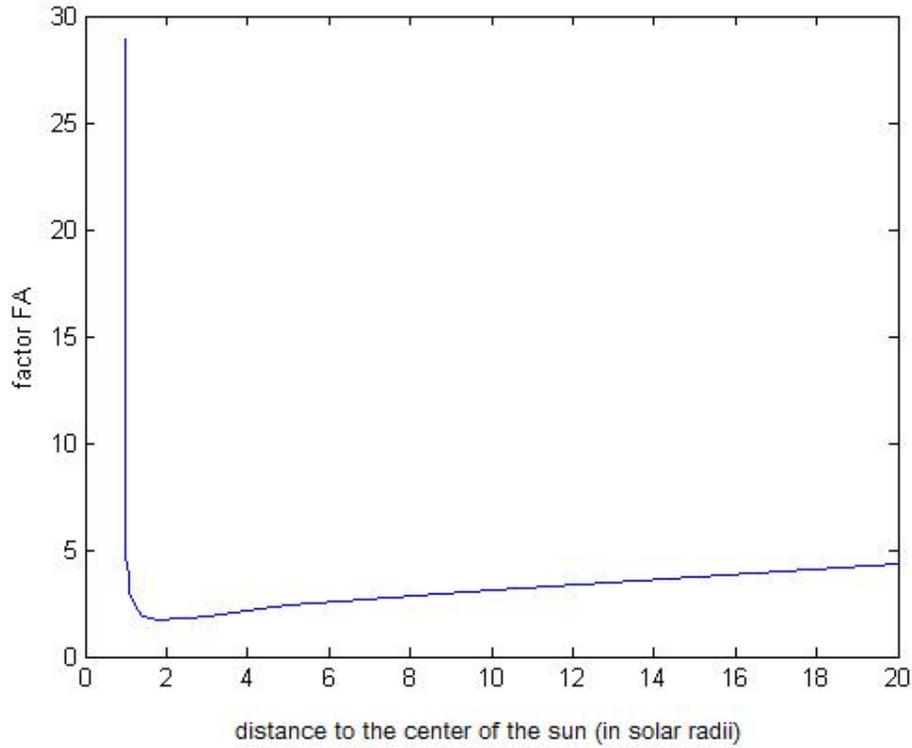

**Fig. 3** The variation of the factor FA$=\dfrac{2e^2}{\pi \hbar}\dfrac{\ln \Lambda}{V_e}$ in the solar atmosphere

For a given apparent distance R, $Z(R)$ can be calculated using formula (7). The theoretical value of the frequency shift caused by the solar corona will be

$$\Delta f(R) = 2292\times 10^6 \times [Z(R)\text{-}Z(9.96R_\odot)] \tag{8}$$

The result calculated is shown in Fig. 4.

The erratic redshift of the observed center frequency relatively to the dashed line in Fig. 2, •f, as a function of the apparent distance R is given in Table 1. The data of Days in 1968, Apparent Distance R and Center frequency shift are copied from table 1 in Merat et al. (1974). The signal of Pioneer 6 disappeared at the apparent distance of approximately 4 solar radii. Fig. 4 gives the erratic drift (asterisks) of the observed center frequency and the curve calculated by formula (8).

**Table 1**. Residual daily shift •f of the center frequency, the apparent distance R from the center of the sun. Distance R and Center frequency shift are copied from table 1 in Merat et al. (1974).

| Days in 1968 | Apparent Distance R (in $R_\odot$) | Center frequency shift (in Hz) | Daily linear drift (dashed line in Fig.2) (in Hz) | Residual shift caused by corona $\Delta f$ (in Hz) |
|---|---|---|---|---|
| Nov. 06 | -9.96 | +374 | +374 | 000 |
| 07 | -9.30 | +355 | +359 | -004 |
| 08 | -8.78 | +332 | +347 | -015 |
| 09 | -8.21 | +297 | +335 | -038 |
| 10 | -7.51 | +285 | +320 | -035 |
| 11 | -7.01 | +253 | +308 | -055 |
| 12 | -6.39 | +238 | +294 | -056 |
| 13 | -5.76 | +229 | +279 | -050 |
| 14 | -5.16 | +209 | +266 | -057 |
| 15 | -4.54 | +185 | +253 | -068 |
| 16 | -4.01 | +168 | +241 | -073 |
| 17 | -3.04 | +144 | +228 | -084 |
| | | | | |
| 29 | +4.20 | -003 | +071 | -074 |
| 30 | +4.86 | -009 | +053 | -062 |
| Dec.,01 | +5.49 | -015 | +039 | -054 |
| 02 | +6.06 | -021 | +027 | -048 |
| 03 | +6.72 | -062 | +013 | -075 |
| 04 | +7.29 | -021 | -000 | -021 |
| 05 | +7.98 | -026 | -013 | -013 |
| 06 | +8.65 | -038 | -027 | -011 |
| 07 | +9.29 | -044 | -039 | -005 |

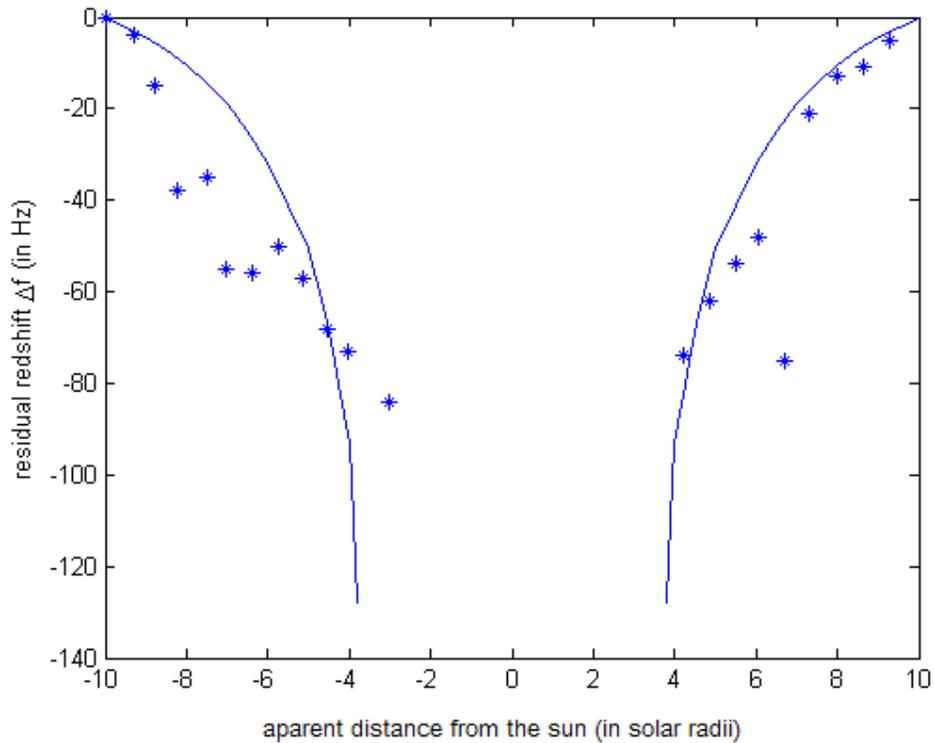

**Fig. 4** The observed erratic drift of the center frequency (*) and the curve calculated by formula (8).
The unit of the apparent distance from the sun is radius of the sun. The unit of the redshift is Hz. The electron density distribution in the solar corona is taken from Allen (1973).

## 4. Discussion

From Fig. 4 it can be seen that the calculated line fits the observed erratic drift of the center frequency, although the fit is not quite well. This is because the experiments by Goldstein were carried out during the solar cycle maximum when a number of solar events were also observed. According to formula (6) the redshift caused by the solar corona is very sensitive to electron density distribution in the solar corona. The solar events caused not only the electron density fluctuation but also the electron density increase related to the quite period after the solar events (caused by the Corona Mass Ejections, CME). This is why most of the observed redshift values are larger than the ones calculated with the model. The 3$^{rd}$ December anomaly, increase in the redshift, was certainly related the solar event. That the signal disappeared abruptly was the result from the fact that the redshift caused by the corona exceeded 100 Hz, the bandwidth used in the observation.

The spectral bandwidth broadening of the signal from Pioneer 6 can be explained by the fluctuation of the electron density along the line of sight. The electron density in the solar corona would be increased after a solar event.

From Fig. 4 it can be seen that the signal disappeared when the apparent distance R was less than 4 solar radii because the erratic redshift exceeded 100 Hz, the bandwidth used in the observation (see Fig. 4 and Table 1.).

The redshift caused by the soft-photon process is very small and very hard to detect, and has long been neglected. However, the soft-photon process has very important significance in astrophysics. The soft-photon process belongs to a reasonable 'tired light' theory.

**Acknowledgement**

I would like to thank Dr. Nailong Wu for the correction and suggestions his made
to greatly improve the English of the manuscript.

**Appendix**

According to Spitzer (1956), in the dilute ionized gas, due to the affects of the other charged particles the velocity of an electron will be changed. In the directions parallel and perpendicular to the original velocity direction of the electron, the changes of velocity can be respectively calculated as follows:

$$\langle (\Delta V_{\rightarrow})^2 \rangle = \frac{A_D}{v} G(l_i V_e); \quad \langle (\Delta V_{\perp})^2 \rangle = \frac{A_D}{v} [\Phi(l_i V_e) - G(l_i V_e)] \tag{a1}$$

where

$$A_D = \frac{8\pi \cdot e^4 z_e z_i n_i \ln \Lambda}{m_e^2}, \qquad \Phi(x) = \frac{2}{\sqrt{\pi}} \int_0^x e^{-y^2} \, dy$$

$$G(x) = \frac{\Phi(x) - x\Phi'(x)}{2x^2}, \qquad l_i = \sqrt{\frac{m_i}{2kT}}$$

$$\Lambda = \begin{cases} \dfrac{3}{2z_e z_i e^3} \left( \dfrac{k^3 T^3}{\pi \cdot n_e} \right)^{\frac{1}{2}} & \text{for } T < 4.2 \times 10^5 \,°K \\ \dfrac{3}{2z_e z_i e^3} \left( \dfrac{k^3 T^3}{\pi \cdot n_e} \right)^{\frac{1}{2}} \left( \dfrac{4.2 \times 10^5}{T} \right)^{\frac{1}{2}} & \text{for } T > 4.2 \times 10^5 \,°K \end{cases}$$

In astronomy the dilute ionized gas consists mainly of ionized hydrogen, and $z_e \approx z_i \approx 1$, $n_i \approx n_e$.

Suppose in the dilute ionized gas, the electrons and protons have the same kinetic temperature. Then $l_i V_e \approx \sqrt{1840}$. During the duration $\Delta t$, due to the interaction between the electron and photons in the propagation process the momentum change of the electron is

$$(\Delta P)^2 = \langle (\Delta P)^2 \rangle \Delta t = \langle (\Delta V_e)^2 \rangle m_e^2 \Delta t \tag{a2}$$

Because

$$\langle (\Delta V_e)^2 \rangle = \langle (\Delta V_\rightarrow)^2 \rangle + \langle (\Delta V_\perp)^2 \rangle \, , \quad \Delta t = n_e^{-\frac{1}{3}} c^{-1}$$

(a2) becomes

$$(\Delta P)^2 \approx \frac{8\pi . e^4 n_e^{\frac{2}{3}} \ln \Lambda}{c V_e} \tag{a3}$$

Taking into account the unit used in Spitzer (1956), the momentum change of the electron in the propagation process is

$$(\Delta P)^2 = \frac{8\pi . e^4 n_e^{\frac{2}{3}} c \ln \Lambda}{V_e} \tag{a4}$$